\documentclass[useAMS,usenatbib,usegraphicx]{mn2e}

\title[The polluted atmosphere of GALEX~J193156.8+011745]
{The heavily polluted atmosphere of the DAZ white dwarf GALEX~J193156.8+011745
\thanks{Based on observations made with ESO telescopes at the La Silla Paranal Observatory
under programmes 83.D-0540 and 283.D-5060.}}
\author[S. Vennes et al.]{S. Vennes$^{1}$\thanks{E-mail: vennes@sunstel.asu.cas.cz}, 
A. Kawka$^{1}$\thanks{E-mail: kawka@sunstel.asu.cas.cz}, and P. N\'emeth$^{2}$\thanks{Email: pnemeth@fit.edu}\\
$^{1}$Astronomick\'y \'ustav AV \v{C}R, Fri\v{c}ova 298,CZ-251 65 Ond\v{r}ejov,
Czech Republic\\
$^{2}$Department of Physics and Space Sciences, 150 W. University Blvd, Florida Institute of Technology, Melbourne, FL 32901, USA
}

\begin{document}

\date{}

\pagerange{\pageref{firstpage}--\pageref{lastpage}} \pubyear{2010}

\maketitle

\label{firstpage}

\begin{abstract}
We report on the discovery of a new heavily polluted white dwarf. 
The DAZ white dwarf GALEX~J193156.8+011745 was identified in a
joint {\it GALEX}/GSC survey of ultraviolet-excess objects.
Optical spectra obtained at ESO NTT show strong absorption lines of
magnesium and silicon and a
detailed abundance analysis based on VLT-Kueyen UVES spectra reveal super-solar
abundances of silicon and magnesium, and near-solar abundances of oxygen, calcium, and iron. 
The overall abundance pattern bears the signature of
ongoing accretion onto the white dwarf atmosphere.
The infrared spectral energy distribution
shows an excess in the H and K bands likely associated with the accretion source.
\end{abstract}


\section{Introduction}

The star GALEX~J193156.8+011745 (hereafter GALEX~J1931+0117) is a hydrogen-rich white dwarf
recently identified in a joint ultraviolet/optical survey \citep{ven2010} 
based  on the {\it Galaxy Evolution Explorer} ({\it GALEX})
all-sky survey\footnote{See \citet{mor2007} for a description of the data products.} and the GSC2.3.2 catalogue. 
The Third U.S. Naval Observatory CCD Astrograph Catalog\footnote{Accessed at VizieR \citep{och2000}.} locates
the star at R.A.(2000)$=19\ 31\ 56.933$, Dec.(2000)$=+01\ 17\ 44.13$ with a proper motion of
$\mu_\alpha\cos{\delta}=-77.6\pm6.4$ and $\mu_\delta=-6.1\pm6.1$ mas~yr$^{-1}$.
The distance modulus implies that the star is nearby ($\approx55$ pc) and that it is imbedded 
in the Galactic plane ($|z|\approx 8$ pc). 
Based on echelle spectroscopy we show that GALEX~J1931+0117 is a DAZ white dwarf and is
one of the most extreme cases of externally polluted white dwarf atmospheres.

The DAZ white dwarf atmospheres are hydrogen-dominated with heavy element abundances ranging from
nearly solar down to detection limits several orders of magnitude below solar \citep{zuc2003,koe2005}.
The DAZ and, more particularly, the DZ phenomena are interpreted as evidence of on-going accretion
of circumstellar material onto white dwarf atmospheres \citep[e.g.,][]{kil2007}. The source of the material has been
ascribed to comets \citep{alc1986} and more recently to tidally-disrupted asteroids \citep[see][]{jur2008}, 
although close, low-mass companions in post-common envelope
binaries are also a known source of material \citep{deb2006,kaw2008}. 

The observational evidence in favour of the accretion of debris material resides in (1) the inferred composition of the accretion flow
\citep{zuc2007} showing an overabundance of refractory elements \citep[see a discussion on the solar system by][]{lod2003}, and (2) the direct spectroscopic signature of gaseous discs 
as in the cases of SDSS~J122859.93+104032.9 and SDSS~J104341.53+085558.2 \citep{gan2006,gan2007}
or dusty discs \citep[e.g.,][]{far2009,bri2009}.

The accreted material may also be supplied by mass loss
from a close, low-mass, and possibly substellar companion. Only a few substellar companion to white dwarf stars are known
such as the white dwarf plus L8-9 resolved pair PHL~5028 \citep{ste2009}
or close DA plus L8 post-common envelope binary WD0137$-$349 \citep{max2006,bur2006}.
Relatively luminous DAZ white dwarfs such as EG~102 \citep{hol1997} may hide low-mass
companions, but
\cite{deb2005} limits them
to substellar types.
The mass loss material captured by the deep potential well of a white dwarf should reflect
the composition of the chromosphere of the red dwarf, which, in most cases, should be close to solar.

In this context, the discovery of the peculiar DAZ white dwarf GALEX~J1931+0117 is timely. 
We present a series of ultraviolet (UV), optical and near infrared (NIR) observations (Section 2) and our model
atmosphere analysis (Section 3.1) revealing a DAZ white dwarf with a peculiar abundance
pattern (Section 3.2). A notable NIR excess (Section 3.3) may
hold clues concerning the nature of this external source. We summarise and conclude in Section 4.

\section{Observations}

\begin{figure}
\includegraphics[width=0.98\columnwidth]{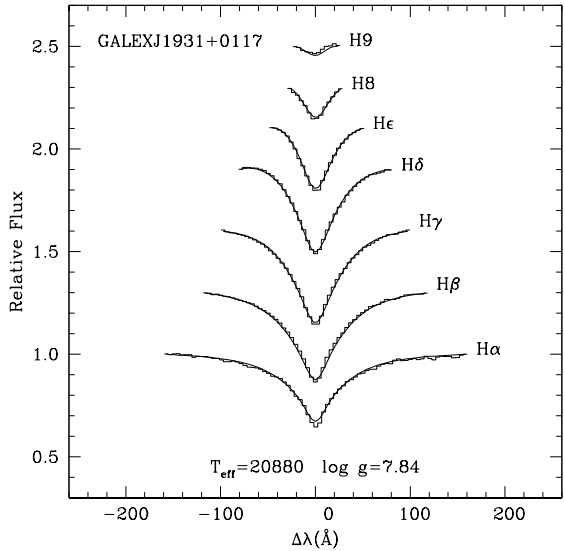}
\caption{Model fits (solid lines) to the low dispersion (EFOSC2) Balmer line profiles (jagged lines) 
and best-fit parameter values ($T_{\rm eff}, \log{g}$). 
}\label{fig1}
\end{figure}

\citet{ven2010} describe a selection of hot sub-luminous stars in a joint {\it GALEX}/GSC2.3.2 survey. 
Briefly,  a UV excess along with measurable proper-motion is interpreted 
as a likely, but not definitive, characteristic of hot subdwarfs (sdB, sdO) or white dwarfs. 
To confirm the selection, follow-up spectroscopy was done for a number of candidates including GALEX~J1931+0117.
We obtained two low-dispersion spectra ($t_{\rm exp}=600$ s) using EFOSC2 attached to the New Technology Telescope (NTT) at 
La Silla Observatory on UT 2009 August 25 (grating \#11, dispersion $=4.15$ \AA\ pixel$^{-1}$). 
The spectral coverage is from 3680 to 7408\AA\ at a resolution of $\approx 15.8$\AA.

Following the identification of GALEX~J1931+0117 as a DAZ white dwarf we obtained
three echelle spectra ($t_{\rm exp}=1450$ s) using UVES attached to the VLT-Kueyen on UT 2009 November 12 (dispersion $\sim$0.02 \AA\ pixel$^{-1}$).
The first two echelle spectra were obtained with the red arm and cover the ranges
5698-7532\AA\ and 7660-9464\AA. A third spectrum obtained with the blue arm covers
the range 3757-4985\AA.

We collected the JHK photometry from the 2MASS survey \citep{skr2006} and 
the {\it GALEX} photometry, corrected for nonlinearity \citep{mor2007}, from the
GR4+5 release. 
Despite a reddening index of $E_{B-V}=0.4054$ \citep{sch1998} in the line of sight toward GALEX~J1931+0117, 
the star is relatively nearby and does not show the effect of reddening ($E_{B-V} \la 0.01$).
Finally, we used the acquisition images (V filter) at NTT to measure the apparent magnitude of
GALEX~J1931+0117.
We employed the acquisition images of the flux calibration standard Feige 110 ($V=11.832$) and EG~21 ($V=11.40$)
obtained on the same night at the NTT to calibrate the measured counts and derive the V magnitudes. 
Table~\ref{tbl-1} summarises the available photometric measurements.
The field surrounding GALEX~J1931+0117 is somewhat crowded with objects
4.8 and 8.3$\arcsec$ away. These separations are sufficient to exclude significant flux contamination
in the optical and infrared images. The ultraviolet (FUV and NUV) images are possibly contaminated. However,
the optical/infrared indices imply that the nearby stars are not significant ultraviolet sources.

\begin{table}
\centering
\caption{Photometry}
\label{tbl-1}
\begin{tabular}{ll}
\hline
Bandpass  & Measurement \\
\hline
{\it GALEX} FUV   & 13.20$\pm$0.10 \\      
{\it GALEX} NUV   & 13.55$\pm$0.10 \\      
 V   & 14.20$\pm$0.02 \\
2MASS J   &  14.66$\pm$0.05 \\
2MASS H   &  14.55$\pm$0.09 \\
2MASS K   &  14.45$\pm$0.10 \\
\hline
\end{tabular} 
\end{table}

\section{Analysis and discussion}

\subsection{Atmospheric parameters}

First, we constrain the effective temperature ($T_{\rm eff}$) and the surface gravity ($\log{g}$).
Two separate measurements show a slight systematic difference in $\log{g}$. 
Fitting the low-dispersion H$\alpha$ to H9 profiles we measure
$T_{\rm eff} =20880\pm240$ and $\log{g}=7.84\pm0.04$, while fitting the high-dispersion H$\beta$ to
H8 we measure $T_{\rm eff} =20890\pm140$ and $\log{g}=7.93\pm0.03$.
Table~\ref{tbl-2} lists the weighted averages. The error bars for the surface gravity were extended in order
to encompass both of our measurements. We computed the corresponding mass, age, and
absolute V magnitude using the mass-radius relations of \citet{ben1999} with
a carbon/oxygen core, thick hydrogen layer ($\log{M_{\rm H}/M_*}=-4$), and zero metallicity.
Finally we estimated the distance using the photometric parallax method.
Figure~\ref{fig1} shows the analysis of the low-dispersion Balmer line profiles.

Next, we calculated the space motion of the star using the
proper-motion (Third U.S. Naval Observatory CCD Astrograph Catalog)
and the barycentric velocity (see Table~\ref{tbl-3})
corrected for the gravitational redshift ($\gamma_g=25.9^{+1.3}_{-2.4}$), 
$v_{\rm bary,corr}=12.5$ km~s$^{-1}$. We employed the methodology described by
\citet{joh1987} and the solar motion from \citet{hog2005}.
The star belongs to the thin disc population.

\begin{table}
\centering
\caption{Properties of the DAZ GALEX~J1931+0117}
\label{tbl-2}
\begin{tabular}{ll}
\hline
Parameter & Measurement \\
\hline
Effective temperature & $T_{\rm eff} = 20890\pm120$ K \\
Surface gravity & $\log{g}=7.90^{+0.03}_{-0.06}$ (c.g.s.) \\
Mass & $M=0.57^{+0.02}_{-0.03}\,M_\odot$   \\
Absolute magnitude & $M_V=10.50^{+0.05}_{-0.10}$ \\       
Cooling age & $t_{\rm cool}=4.5\pm0.6\times10^7$ yrs \\
   &   \\
Distance & $d=55^{+5}_{-3}$ pc \\
Kinematics & $(U,V,W)=(28,5,22)$ km~s$^{-1}$ \\
\hline
\end{tabular} \\
\end{table}

Our basic model atmosphere grid was computed in local thermodynamic equilibrium (LTE),
but additional non-LTE model atmospheres were computed using TLUSTY \citep{hub1995,lan1995}.
We determined the non-LTE metallicity vector by fitting a non-LTE model at 20900 K and $\log{g}=7.9$ to a grid
of pure-hydrogen LTE models. We measured $\Delta T_{\rm eff}=+120$ K and $\Delta \log{g}=-0.04$. 
We conclude that despite the high-metallicity of the atmosphere, the systematic
error introduced by using pure-H LTE models is within statistical errors. 

\subsection{Photospheric abundances and diffusion}

\begin{figure}
\includegraphics[width=0.98\columnwidth]{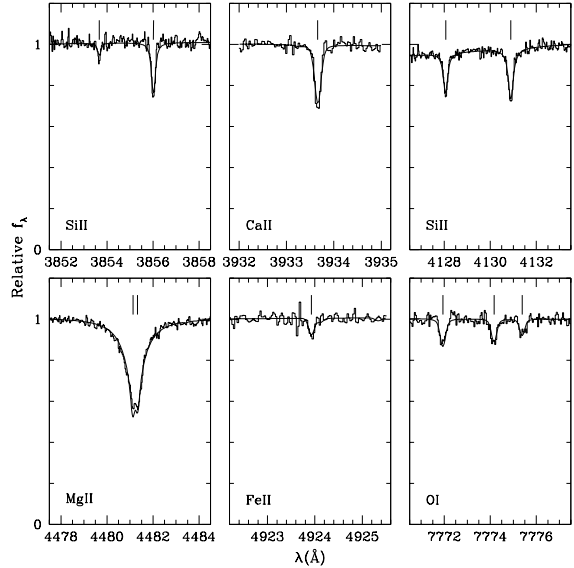}
\caption{Echelle spectra (UVES) of the DAZ GALEX~J1931+0117 showing a strong
Mg doublet and weaker oxygen, silicon, calcium and iron lines.}\label{fig2}
\end{figure}

Table~\ref{tbl-3} lists the heavy-element lines identified in the UVES spectra. The average barycentric velocity of
the trace elements ($v_{\rm bary}=  38.4\pm3.2$ km~s$^{-1}$) is in agreement with the H$\alpha$ 
velocity $v_{\rm H\alpha} = 37.6\pm0.8$ km\,s$^{-1}$.
We determined the photospheric abundances by fitting the observed line profiles with non-LTE models.
Figure~\ref{fig2} shows selected heavy element line profiles and best-fit models. We did not notice 
line splitting due to a magnetic field at the resolution limit ($R=\lambda/\Delta\lambda \approx 40,000$).
Since $\Delta\lambda/\lambda = 4.7\times10^{-7}\lambda\,B$ where $\lambda$ is in \AA\ and $B$ in MG, the limit implied is
$B\la 10$ kG.

\begin{table}
\centering
\caption{UVES line identification}
\label{tbl-3}
\begin{tabular}{cccc}
\hline
 Ion & $\lambda$ & E.W. & $v_{\rm bary}\,^a$   \\
      &    (\AA) & (m\AA) &  (km~s$^{-1}$) \\
\hline
 \mbox{Si\,{\sc ii}} & 3853.660 & 17.  & 36.6 \\
 \mbox{Si\,{\sc ii}} & 3856.020 & 56.  & 36.8\\
 \mbox{Si\,{\sc ii}} & 3862.609 & 50.  &36.7 \\
 \mbox{Ca\,{\sc ii}} & 3933.663 & 50.  & 37.1\\ 
 \mbox{Ca\,{\sc ii}} & 3968.469 & 11.  & 37.7\\ 
 \mbox{Si\,{\sc ii}} & 4128.070 & 56.  & 36.4\\
 \mbox{Si\,{\sc ii}} & 4130.890 & 83.  & 36.2\\
 \mbox{Mg\,{\sc ii}} & 4481.13/.33 & 579.  & 33.7 \\
 \mbox{Fe\,{\sc ii}}  & 4549.474 &  6.  & 35.3 \\
 \mbox{Si\,{\sc iii}} & 4552.622 & 11.  & 38.1\\
 \mbox{Fe\,{\sc ii}}  & 4923.927 & 12.  & 37.6\\
 \mbox{Si\,{\sc ii}} & 5957.560 & 34.  &  40.8\\
 \mbox{Si\,{\sc ii}} & 5978.930 & 49.  &  43.2\\
 \mbox{Si\,{\sc ii}} & 6347.100 & 132.  & 36.4\\
 \mbox{Si\,{\sc ii}} & 6371.360 & 102.  & 36.4\\
 \mbox{O\,{\sc i}}   & 7771.944 & 39.  & 39.2 \\ 
 \mbox{O\,{\sc i}}   & 7774.166 & 31.  & 39.0  \\
 \mbox{O\,{\sc i}}   & 7775.388 & 22.  & 41.5 \\
 \mbox{Mg\,{\sc ii}} & 7877.054 & 46.  & 45.5 \\
 \mbox{Mg\,{\sc ii}} & 7896.366 & 103.  & 45.5\\
 \mbox{Ca\,{\sc ii}} & 8542.091 & 18.  &  35.8\\
\hline
\end{tabular} \\
$^a$ Average barycentric velocity $v_{\rm bary}=  38.4\pm3.2$ km~s$^{-1}$.
\end{table}

Table~\ref{tbl-4} lists the measured abundances. Helium and carbon are not detected and we employed
\mbox{He\,{\sc i}}$\lambda$5875.7 and \mbox{C\,{\sc ii}}$\lambda$4267.001-183 to set abundance upper limits. 
The infrared calcium triplet region shows weak photospheric absorption but no gaseous disc emission with a limit of
$E.W.\le 1$\AA. The abundance of magnesium in GALEX~J1931+0117 exceeds that of SDSS~J122859.93+104032.9
by a factor of 5 but no evidence of a gaseous disc is found in disagreement with a correlation
between magnesium abundance and calcium emission strength in hot DAZ white dwarfs found by \citet{gan2007}.
However, \citet{jur2008} argues that gaseous discs are rare and that in the two cases discussed by \citet{gan2007}
an additional source of heating must be responsible for this unusual phenomenon.

The near-solar heavy element abundance in the atmosphere of GALEX~J1931+0117 must be 
supplied from an external source. The accreted
material reaches optical depth unity, $\tau_R\approx 1$, and, assuming near steady-state, 
appears in the optical spectra at the measured abundances, $n({\rm X})/n({\rm H})$.
The elements are supplied in the accretion flow in proportions $n({\rm X})/n({\rm H})_{\rm flow}$ that are not
necessarily solar. The steady-state abundances are proportional to (1) the supplied abundances and (2)
their respective diffusion time-scales. For example, the measured abundance of element X relative to silicon is given by
\begin{equation}
\frac{n({\rm X})}{n({\rm Si})} =\frac{\tau({\rm X})}{\tau({\rm Si})}\,\frac{n({\rm X})}{n({\rm Si})}_{\rm flow}
\end{equation}
where the time-scale $\tau({\rm X})$ is given at a selected point in the atmosphere by
\begin{equation}
\tau({\rm X}) = \frac{m}{\rho\,v}
\end{equation}
where $m$ is the column mass density (g\,cm$^{-2}$), $\rho$ is the local mass density (g\,cm$^{-3}$), and $v$ is the diffusion velocity (cm\,s$^{-1}$) at that point.
These quantities are calculated at $\tau_R=1$.
Both $m$ and $\rho$ are provided by the model atmosphere. The diffusion velocity $v=v(T,n_p,X_i)$
is calculated following the prescription of \citet{fon1979} in a fully ionized hydrogen gas ($Z_{\rm H}=1$),
and the remaining variables, temperature $T$, proton density $n_p$, and the ionization fractions $X_i$ of element $X$ 
are also provided by the model atmosphere.  
The diffusion velocity is averaged over all ionization species following \citet{mon1976}.
The relatively high concentrations of neutral helium and oxygen contribute toward 
shorter diffusion time scales compared to elements with lower concentrations of
neutral species.

Table~\ref{tbl-4} summarises our calculations. Using Equation (1) we calculated the abundances 
relative to silicon
that are required in the accretion flow to reproduce the observed abundance ratios and
compared them to solar ratios.

Again, assuming steady state accretion,
the total mass of element X accreted per year and mixed into an atmosphere of mass $M=4\pi R^2\,m$, where
$R$ is the white dwarf radius, is given by
\begin{equation}    
\dot{M} = A({\rm X})\frac{n({\rm X})}{n({\rm H})} \frac{4\pi R^2\,m}{\tau({\rm X})} 
\end{equation}    
where $A$(X) is the atomic weight of element X and the abundance is given relative to
the main constituent. For example, magnesium and silicon must be accreted at a rate of $4\times10^{15}$
and $6\times10^{15}$ g\,yr$^{-1}$, respectively. The total accretion rate for magnesium and silicon
is well within the range predicted by the tidal destruction of asteroids 
\citep{jur2008},
but the relatively low abundance of calcium argues against this scenario.
Abundance measurements in the dusty helium-rich white dwarfs GD~40 \citep{kle2010} and GD~362 \citep{koe2009},
and hydrogen-rich white dwarf G~29-38 \citep{koe2009} 
imply a calcium-rich accretion flow with $n({\rm Ca})/n({\rm Mg})\sim 0.1-1$. The accretion flow
in GALEX~J1931+0117 is low in calcium ($n({\rm Ca})/n({\rm Mg})\sim 0.02$), somewhat below
the solar ratio. Within uncertainties the accretion flow has near solar composition.

Alternatively, assuming that the accreted material is supplied by mass loss from
a low-mass companion, the accreted mass of the main constituent (hydrogen) is scaled
on the accreted mass of trace elements. Averaging the observed variations between elements,
the total accretion rate is estimated at $\dot{M}_{\rm acc}=3\times10^{-14}\,M_\odot$\,yr$^{-1}$.
Using the post-common envelope white dwarf plus brown dwarf binary WD0137$-$349 as an example
we arbitrarily locate a L5 companion (see section 3.3) of $0.05\,M_\odot$ \citep[see][]{kir2005}
at a separation of $a=1\,R_\odot$. 
The fraction of material captured by the white dwarf
would then dictate the required mass-loss rate for the companion. Assuming in turn
a geometrical cross-section, a gravitationally-focused cross-section, or fluid accretion 
\citep{wes1984,deb2006}, the fraction is estimated at $5\times10^{-5},\ 3\times10^{-3}$,
or 0.02-0.08, respectively. Under the assumption of fluid accretion, the required 
mass-loss rate is $\dot{M}_{\rm acc}=0.4-1.5\times10^{-12}\,M_\odot$\,yr$^{-1}$.
This rate exceeds the solar rate or the upper limit derived for
Proxima Centauri implying that mass loss may be stimulated by
tidal or irradiation effects in a close binary \citep[see a discussion by][]{deb2006}.

\begin{table*}
\centering
\begin{minipage}{160mm}
\caption{Photospheric abundances}
\label{tbl-4}
\begin{tabular}{lllllll}
\hline
X  & $n$(X)$/n$(H) & $n$(X)$/n$(Si) & $n$(X)$/n$(Si)$_\odot$ $^{a}$ & $\tau$(Si)$/\tau$(X) $^{b}$ & $n$(X)$/n$(Si)$_{\rm flow}$ $^{c}$ & Departure from solar$^{d}$ \\
\hline
He  & $<5\times10^{-4}$           &  $<7.8$          & $2.6\times10^3$ & 60.  & $\la 5\times10^2$  & $\la0.2$ \\
C   &  $<7\times10^{-5}$          &  $<1.1$          & $7.6$           & 1.07 & $\la 1.2$  & $\la 0.2$  \\
O   & $2.4\pm0.2\times10^{-4}$    &  $3.8\pm0.4$     & $14.$           & 7.4  & $\sim30$   & $\sim2$   \\
Mg  & $7.1\pm0.3\times10^{-5}$    &  $1.1\pm0.1$     & $1.1$           & 0.81 & $\sim0.9$  & $\sim0.8$   \\
Si  & $6.4\pm0.3\times10^{-5}$    &    ...           & ...             & ...  & ...        & ...        \\
Ca  & $9.0\pm1.8\times10^{-7}$    &  $0.014\pm0.003$ & $0.063$         & 1.23 & $\sim0.02$ & $\sim0.3$ \\
Fe  & $2.9\pm0.6\times10^{-5}$    &  $0.45\pm0.09$   & $0.87$          & 1.86 & $\sim0.8$ & $\sim1$   \\
\hline
\end{tabular} \\
$^a$ Abundances in the solar photosphere from \citet{gre2007}.
$^b$ Ratio of diffusion time scales at $\tau_R=1$.
$^c$ Predicted abundance ratio in the accretion flow.
$^d$ Predicted abundance ratio in the accretion flow relative to solar.
\end{minipage}
\end{table*}

\subsection{Infrared excess: low-mass star or dusty disc}

\begin{figure}
\includegraphics[width=0.98\columnwidth]{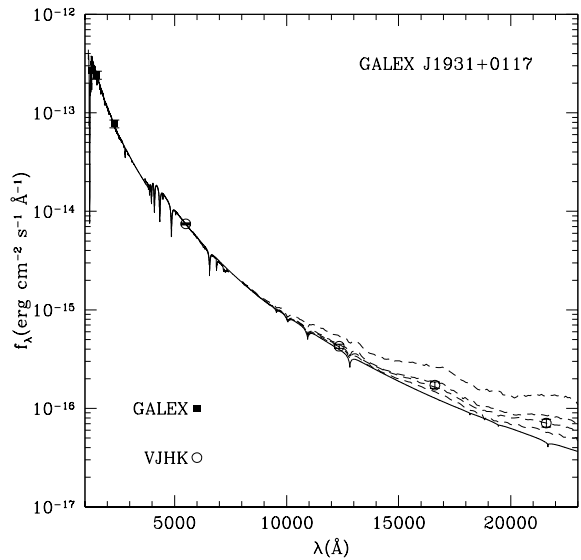}
\caption{UV to NIR spectral energy distribution 
based on {\it GALEX} FUV and NUV photometry, the NTT V magnitude, and
2MASS JHK photometry compared to a white dwarf model (full line). 
Contributions from (top to bottom dashed lines) a L2, L4, L5, and L6 dwarf are added
to the white dwarf model.}\label{fig3}
\end{figure}

Figure~\ref{fig3} compares the spectral energy distribution from the UV to the NIR to computed
composite spectra. 
Adopting class averages of $M_J=12.0, 13.0, 13.5,$ and 14.0 for L2, L4, L5, and L6, respectively \citep{kir2005}, we scaled
corresponding L templates from the IAC M-L-T Dwarf Catalog \citep{mar2005} to a distance of 55 pc, and added the flux to the projected NIR continuum of the DAZ white
dwarf.
The measured excess would correspond to a L5 dwarf with 
$T_{\rm eff}\sim1700$ K \citep{kir2005}.

Assuming $M_2=0.05\,M_\odot$ and $a=1\,R_\odot$,
the predicted L5 and DAZ velocity semi-amplitudes are $K_{\rm L5} = 314\cdot\sin{i}$ km\,s$^{-1}$ and $K_{\rm DAZ} = 28\cdot\sin{i}$
km\,s$^{-1}$, respectively.
Radial velocity shifts in the white dwarf spectra may be detectable in future observations. 
The presence of a close L dwarf companion should lead to notable optical/infrared light variations. 
In particular, variable H$\alpha$ emission is a noted spectroscopic signature of post-common envelope
systems \citep[e.g., WD0137$-$349,][]{max2006}, but it is not apparent in any of our spectra.

We cannot exclude the presence of a dusty circumstellar environment 
in addition to a L dwarf companion. 
However, by showing a steeper 
decline over the H- and K-bands the NIR spectral energy distribution of GALEX~J1931+0117
appears different than that of the warm, dusty white dwarf SDSS~J122859.93+104032.9 \citep{bri2009}.
If present, dust should reside
at a distance from the star where the equilibrium temperature between dust and stellar radiation is below
the sublimation temperature, $T_{\rm sub}$ \citep[see][]{jur2008}. For $T_{\rm sub}$ between 1350 and 1650 K \citep[see][]{lod2003}
the minimum distance from GALEX~J1931+0117 or SDSS~J122859.93+104032.9 is between 1.7 and 1.1 $R_\odot$. 
In fact, \citet{bri2009} found that the gaseous and dusty discs in SDSS~J122859.93+104032.9
co-exist at a distance $\la 1.2\,R_\odot$ from the white dwarf.
Although GALEX~J1931+0117 and
SDSS~J122859.93+104032.9 have nearly identical luminosities, the lack of a gaseous disc in the former and the dissimilarity
of their NIR spectral energy distributions imply different environments.

\section{Summary and conclusions}

We identified and analysed the properties of one of the most heavily polluted white dwarfs known.
The source of the material remains unknown, although the measured NIR excess may be attributed to
a L5 dwarf or to a warm debris disc.
The measured abundances suggest that the material is accreted onto the atmosphere in solar 
proportions and 
favour a model involving a L dwarf companion in a close orbit. However, 
the H$\alpha$ emission notable in such systems is absent in GALEX~J1931+0117 and the
mass loss imposed on the companion appears excessive.
NIR intermediate dispersion spectroscopy should help determine the nature of the accretion source.
Additional optical spectra will be useful to trace putative orbital motions while
Space Telescope Imaging Spectrograph high-dispersion spectra will be useful to constrain further 
the carbon abundance and extend the pattern to less abundant elements.

\section*{Acknowledgments}
S.V. and A.K. are supported by GA AV grant numbers IAA300030908 and IAA301630901, respectively, and by GA \v{C}R grant number P209/10/0967.
A.K. also acknowledges support from the Centre for Theoretical Astrophysics (LC06014). 
We thank the referee for useful suggestions.
This publication makes use of data products from the Two Micron All Sky Survey, which is a joint 
project of the University of Massachusetts and the Infrared Processing and Analysis Center/California 
Institute of Technology, funded by the National Aeronautics and Space Administration and the National 
Science Foundation.

\label{lastpage}

\end{document}